\documentclass [12 pt]{article}
\def\be{\begin{equation}}
\def\eea{\end{eqnarray}}
\def\bea{\begin{eqnarray}}
\def\ee{\end{equation}}

\author{F. Kheirandish$^{1}$ \footnote{fardin$_{-}$kh@phys.ui.ac.ir} and M.
Amooshahi$^{1}$ \footnote{amooshahi@sci.ui.ac.ir}
\\ $^{1}$ {\small Department of Physics, University of Isfahan,}
\\ {\small Hezar Jarib Ave., Isfahan, Iran.}}
\title{Dissipation in Quantum Mechanics, Scalar and Vector Field Theory}
\begin{document}
\maketitle
\begin{abstract}
A new minimal coupling method is introduced. A general dissipative
quantum system is investigated consistently and systematically.
Some coupling functions describing the interaction between the
system and the environment are introduced. Based on coupling
functions, some susceptibility functions are attributed to the
environment explecitly. Transition probabilities relating the way
energy flows from the system to the environment are calculated
and the energy conservation is explecitly examined. This new
formalism is generalized to the dissipative scalar and vector
field theories along the ideas developed for the quantum dissipative systems.

\noindent
\end{abstract}
\section{Introduction}
There are basically two approaches to study dissipative quantum
systems. One is found in the interactions between two systems via
an irreversible energy flow \cite{I1,I2}. The second approach is
a phenomenological treatment under the assumption of
nonconservative forces \cite{I3,I4}. In studying nonconservative
systems, it is essential to introduce a time dependent
Hamiltonian which describes the damped motion. Such a
phenomenological approach for the study of dissipative quantum
systems, especially a damped harmonic oscillator, has a rather
long history. Caldirola and Kanai \cite{I5,I6} adopted the
Hamiltonian
\begin{equation}\label{dI1}
H(t)=e^{-\beta t}\frac{p^2}{2m}+e^{\beta t}\frac{1}{2}m\omega_0^2
q^2,
\end{equation}
which leads exactly to the classical equation of motion of a
damped harmonic oscillator
\begin{equation}\label{dI2}
\ddot{q}+\beta\dot{q}+\omega_0^2q=0.
\end{equation}
The quantum aspect of this model has been studied in a great
amount of literature. In those studies some peculiarities of this
model and some features of it have appeared to be ambiguous
\cite{I7}-\cite{I15}. There are significant difficulties in
obtaining the quantum mechanical solutions for the Caldirola
-Kanai Hamiltonian. Quantization with this Hamiltonian violates
the uncertainty relations, i.e., the uncertainty relations vanish
as time goes to infinity \cite{I16}-\cite{I19}.

 Based on Caldirola-kanai
 Hamiltonian, it has been constructed equivalent theories by
 performing a quantum canonical transformation and has been used
 the path integral techniques to calculate the exact propagators of
 such theories, also the time evolution of given initial wave
 functions, have been studied using the obtained propagators \cite
 {I20}.

  In the framework of the phenomenological approach Lopez and
 Gonzalez \cite{I21} have taken the external non conservative forces that has
 linear and quadratic dependence with respect to velocity. They
 have deduced classical constants of motion and Hamiltonian
 for these systems and eigenvalues of these constants  have been
 quantized through perturbation theory.

  A simple pseudo-Hamiltonian formulation has been proposed by
 Kupriyanov and et al \cite{I22}. Starting from this pseudo-
 Hamiltonian formulation a consistent deformation quantization has
 been developed that involve  a non-stationary star product and an extended
 operator of time derivative differentiating the star
 product. A complete consistent quantum-mechanical
 description for any linear dynamical system with or without
 dissipation has been constructed in this scheme.\\
 In another approach to quantum dissipative systems, one tries to bring about
 the dissipation as a results of an averaging over all the coordinates of the bath system,
 where one considers the whole system as composed of two parts, our main system and
 the bath system which interacts with the main system and causes the
 dissipation of energy on it \cite{I23}-\cite{I30}.\\
 The macroscopic description of a quantum particle with passive
 dissipation and moving in an external potential $v(\vec{x})$ is
 formulated in terms of Langevin-Schr\"{o}dinger equation
 \cite{M1,M2,M3}
 \begin{equation}\label{d.1}
 m\ddot{x}+\int_0^t dt'
 \mu(t-t')\dot{\vec{x}}(t')=-\vec{\nabla}v(\vec{x})+F_N(t).
 \end{equation}

 The coupling with the heat-bath in microscopic levels corresponds
 to two terms in macroscopic description. A mean force characterized
 by a memory function $\mu(t)$ and an operator valued random
 force $ F_N(t)$. These two terms have a fluctuation-dissipation
 relation and both are required for a consistent quantum
 mechanical description of the particle. In \cite{M3}, there are
 some models for interaction of the main system with the heat-bath which lead to
the macroscopic Langevin-Schr\"{o}dinger equation.

 The layout of the paper is as follows:
In section 2, we consider a particle moving in a one-dimensional
external potential and take the absorptive environment of the
particle as a Klein-Gordon field which interact with the momentum
of the particle through a minimal coupling term. In this approach,
 the Langevin-Schrodinger equation(\ref{d.1}) is obtained as the
macroscopic equation of motion of the particle and the noise force
$F_N(t)$, is derived in terms of the coupling function and the
ladder operators of the environment. By choosing a special form
for the coupling function, a friction force proportional to the
velocity of the particle is obtained . For an initially exited one
dimensional damped harmonic oscillator, it is shown that the
entire energy of the oscillator will be absorbed by its
environment.
 In section 3, we are concerned with a particle moving
in three-dimensional absorptive environment. In this section we
model the environment by two independent Klein-Gordon fields $B$
and $\tilde{B}$. The field $B$ interacts with the momentum of the
particle through a minimal coupling term and the field
$\tilde{B}$, interacts with the position operator of the particle
similar to a dipole interaction term. A generalized
Langevin-Schrodinger equation is obtained which contains two
memory functions describing the absorption of energy of the
particle by its environment. As an simple example, the spontaneous
decay constant and the shift frequency of a two-level system,
embedded in an absorptive environment, is calculated.

In sections 4 and 5, the ideas set up in sections 2 and 3, are
applied to quantum dissipative systems described by scalar or
vector fields.
 \section{One-dimensional quantum dissipative systems}
\subsection{Quantum dynamics}
Quantum mechanics of a one-dimensional damped system can be
investigated in a systematic and consistent way by modeling the
environment of the system with a quantum field $B$ which
interacts with the system through a {\bf minimal coupling term}.
For this purpose, let the damped system be a particle with mass
$m$, under the influence of an external potential $v(q)$. We take
the total Hamiltonian of the system and it's environment as
\cite{M1}
\begin{equation} \label{d1.1}
H=\frac{(p-R)^2}{2m}+v(q)+H_B,
\end{equation}
where $q$ and $p$ are position and canonical conjugate momentum
operators of the particle respectively and satisfy the canonical
commutation rule
\begin{equation}\label{d1.2}
[q,p]=i\hbar.
\end{equation}
 In Hamiltonian (\ref{d1.1}), $ H_B $ is the Hamiltonian of the environment. Now we model the
 environment by a massless Klein-Gordon field, such that we can write the environment
 Hamiltonian as
 \begin{equation}\label{d1.3}
H_B(t)=\int_{-\infty}^{+\infty}d^3k  \hbar\omega_{\vec{k}}
b_{\vec{k}}^\dag(t) b_{\vec{k}}(t),
\end{equation}

where $\omega_{\vec{k}}$, is the dispersion relation of the
environment. The annihilation and creation operators $
b_{\vec{k}}$, $b_{\vec{k}}^\dag $, in any instant of time,
satisfy the following commutation relations
\begin{equation}\label{d1.4}
[b_{\vec{k}}(t),b_{\vec{k}'}^\dag(t)]=\delta(\vec{k}-\vec{k}').
\end{equation}
  Operator $R$ in relation (\ref{d1.1}),
has the basic role in interaction between the system and it's
environment and is defined by
\begin{equation}\label{d1.5}
R(t)=\int_{-\infty}^{+\infty}d^3k [f(\omega_{\vec{k}})
b_{\vec{k}}(t)+f^*(\omega_{\vec{k}})b_{\vec{k}}^\dag(t)],
\end{equation}
where the function $f(\omega_{\vec{k}})$ is called the coupling
function between the particle and its environment. In definition
of operator $R$, the following physical assumptions are considered:\\
\\
1. Since the total Hamiltonian (\ref{d1.1}) is Hermitian, so the operator $R$
should be a Hermitian operator.\\
\\
2. We are interested in linear dissipative systems and for such
systems, the operator $R$ should be a linear combination of
annihilation and creation operators $b_{\vec{k}}$ and $b_{\vec{k}}^\dag$.\\
\\
3. Operator $R$, should depend on macroscopic characters of the
environment, because we are interested in macroscopic equations of
motion of the particle. It is remarkable to note that the
macroscopic characters of the environment are reflected in its
macroscopic susceptibility against the motion of the particle. In
the following, we see that the susceptibility, is related to both
 the dispersion relation $\omega_{\vec{k}}$ and the coupling
function $f(\omega_{\vec{k}})$ and these considerations are
contained in definition of the operator $R$.

 The Heisenberg
equations for the position and momentum of the particle are
\begin{eqnarray}\label{d1.6}
&&\dot{q}=\frac{\imath}{\hbar}[H,q]=\frac{p-R}{m},\nonumber\\
&&\dot{p}=\frac{\imath}{\hbar}[H,p]=-\frac{\partial v}{\partial
q},
\end{eqnarray}
 where $H$ is the total Hamiltonian (\ref{d1.1}). By eliminating $p$ from relations
 (\ref{d1.6}), we obtain
\begin{equation}\label{d1.7}
m\ddot{q}=-\frac{\partial v}{\partial q}-\dot{R}.
\end{equation}
Also using (\ref{d1.4}), the Heisenberg equation for $
b_{\vec{k}}$, is
\begin{equation}\label{d1.8}
\dot{b}_{\vec{k}}=\frac{\imath}{\hbar}[H,b_{\vec{k}}]=-i\omega_{\vec{k}}
b_{\vec{k}}+\frac{\imath}{\hbar}\dot{q}f^*(\omega_{\vec{k}}),
\end{equation}
which formally can be solved as
\begin{equation}\label{d1.9}
b_{\vec{k}}(t)=b_{\vec{k}}(0)e^{-i\omega_{\vec{k}}
t}+\frac{\imath}{\hbar}f^*(\omega_{\vec{k}}) \int_0^t d t'
e^{-i\omega_{\vec{k}}(t-t')} \dot{q}(t').
\end{equation}
 Now substituting $ b_{\vec{k}}(t)$ from this equation into
(\ref{d1.5}), we obtain
\begin{equation}\label{d1.10}
R(t)=R_N(t)+\int_0^{|t|} dt'\chi(|t|-t')\dot{q}(\pm t'),
\end{equation}
where the upper(lower) sign corresponds to $t>0$$(t<0)$,
respectively. Inspired by electrodynamics in a media, let us call
the memory function
\begin{eqnarray}\label{d1.11}
&&\chi(t)=\frac{8\pi}{\hbar }\int_0^\infty d |\vec{k}|
|\vec{k}|^2|f(\omega_{\vec{k}})|^2\sin\omega_{\vec{k}} t \hspace{1cm}t>0\nonumber\\
&&\chi(t)=0\hspace{6.2500cm}t\leq0
\end{eqnarray}
as the susceptibility of the environment.  The operator $R_N$
\begin{equation}\label{d1.12}
R_N(t)=\int_{-\infty}^{+\infty} d^3 k
[f(\omega_{\vec{k}})b_{\vec{k}}(0)
e^{-i\omega_{\vec{k}}t}+f^*(\omega_{\vec{k}})b_{\vec{k}}^\dag(0)e^{i\omega_{\vec{k}}t}],
\end{equation}
is a noise operator.

 The noise forces are necessary for dealing
with dissipative quantum systems in a consistent way, without
theses noises, some fundamental postulates of quantum mechanics
like canonical commutation relations or Heisenberg uncertainty
relations usually are violated.

 The form of the equation (\ref{d1.10}) suggests that
take this equation as a constitutive equation of the environment
and interpret the operator $R$ as some kind of the polarization
of the environment, induced by the particle. A feature of this
approach is its flexibility to choosing an appropriate dispersion
relation $\omega_{\vec{k}}$ and finding the corresponding
coupling function for a given susceptibility. In fact, in
definition (\ref{d1.11}) of the susceptibility, there are two
main functions, namely the coupling function $f(\omega_{{k}})$
and the dispersion relation $\omega_{{k}}$, if one takes the
simplest assumption $\omega_{{k}}=c|\vec{k}|$ for the dispersion
relation and write (\ref{d1.11}) as
\begin{eqnarray}\label{d1.11.1}
&&\chi(t)=\frac{8\pi}{\hbar c^3}\int_0^\infty d\omega
\omega^2|f(\omega)|^2\sin\omega t, \hspace{1.25 cm} t>0, \nonumber\\
 &&\chi(t)= 0, \hspace{5.75 cm} t\leq0,
\end{eqnarray}
then by inverting (\ref{d1.11.1}), the corresponding coupling
function for a given susceptibility, can be obtained as
\begin{eqnarray}\label{d1.13}
&&|f(\omega)|^2= \frac{\hbar c^3 }{4\pi^2\omega^2}\int_0^\infty
dt\chi(t)
\sin\omega t, \hspace{1.25 cm} \omega>0 ,\nonumber\\
&&|f(\omega)|^2=0,\hspace{5.25 cm}  \omega=0.
\end{eqnarray}
 Other choices for the dispersion relation, just lead to a
redefinition of the coupling function and also more difficult
mathematical expressions which basically may not allow to take an
inverse of (\ref{d1.11}) and find the corresponding coupling
function in terms of the susceptibility $\chi(t)$. It is
remarkable to note that one can take the real dispersion relation
of the medium and enrich the model in this way but in this
approach this is only a choice. From now on, we take a linear
dispersion relation as $\omega_{\vec{k}}=c|\vec{k}|$.

Substituting $R$ from (\ref{d1.10}) into (\ref{d1.7}), one can
show easily that the equation of motion for $ q $ is
\begin{equation}\label{d1.14}
m\ddot{q}\pm\frac{d}{dt}\int_0^{|t|} d t'\chi(|t|-t')\dot{q}(\pm
t')=-\frac{\partial v}{\partial q}-\dot{R}_N,
\end{equation}
this is the macroscopic Langevin-Schr\"{o}dinger equation for a
dissipative system \cite{M3}. If we take the
Langevin-Schr\"{o}dinger equation as the basic macroscopic
equation of quantum dissipative systems, then the microscopic
interaction of the particle with its environment can be
macroscopically formulated by modeling the environment with a
quantum field $B$ and using the Hamiltonian (\ref{d1.1}), which
finally leads to the correct macroscopic Langevin-Schr\"{o}dinger
equation.

 In fact, the present approach can be considered as a
generalization of the Caldeira-Legget model \cite{I25,I26} , where
for dissipative quantum systems, they model the environment of the
system by a collection of harmonic oscillators such that the
position operator of the main system, is coupled with all of the
position operators of the environment oscillators. While the
properties of the environment may in some cases be chosen on the
basis of a microscopic model, this does not have to be the
Caldeira-Legget model. As an example we mention an Ohmic resistor
which as a linear electric element should be well described by
the Caldeira-Leggett model. On the other hand the underlying
mechanism leading to dissipation in a resistor may be more
complicated than that implied by the model of a collection of
harmonic oscillators. Modeling the environment with a collection
of Harmonic oscillators, with a suitable interaction, means that
an averaging over complicated microscopic interactions of the
particle with its environment is done and the environment is
effectively equivalent to this modeling.
\subsection{Quantum damped harmonic oscillator}
 For a one dimensional
harmonic oscillator with mass $m$ and frequency $\omega_0$, we
have $
 v(q)=\frac{1}{2}m\omega_0^2q^2 $ and therefore we can
 write (\ref{d1.14}) as
 \begin{equation}\label{d1.15}
m\ddot{q}+m\omega^2_0q\pm\frac{d}{dt}\int_0^{|t|} d
t'\chi(|t|-t')\dot{q}(\pm t')=-\dot{R}_N.
\end{equation}
This equation can be solved for negative and positive times using
the Laplace transformation technique. For any time dependent
operator $g(t)$, the forward and backward Laplace transformations
are by definition
\begin{equation}\label{d1.16}
\underline{g}^f(s)=\int_0^\infty dt g(t)e^{-st},
\end{equation}
 and
\begin{equation}\label{d1.17}
\underline{g}^b(s)=\int_0^\infty dt g(-t)e^{-st},
\end{equation}
respectively. Let $\underline{\chi}(s)$ be the Laplace
transformation of $\chi(t)$, then taking the Laplace transform of
the equation (\ref{d1.15}), we obtain the forward and backward
Laplace transformation of $q(t)$, i.e., $\underline{q}^{f,
b}(s)$, as
\begin{eqnarray}\label{d1.18}
&&\underline{q}^{f,b}(s)=\frac{ms+s\underline{\chi}(s)}{m\omega_0^2
+ms^2+s^2\underline{\chi}(s)}q(0)\pm\frac{m}{m\omega_0^2+ m
s^2+s^2\underline{\chi}(s)}\dot{q}(0)+\nonumber\\
&&\frac{\mp s\underline{R}^{f,b}_N(s)\pm R_N(0)}{m\omega_0^2+ m
s^2+s^2\underline{\chi}(s)},
\end{eqnarray}
where the upper (lower) sign corresponds to forward (backward)
Laplace transformation respectively.
\subsubsection{A special case}
Let us as a special case, take the coupling function $f(\omega)$
as
\begin{equation}\label{d1.19}
|f(\omega)|^2=\frac{\beta\hbar c^3}{4\pi^2\omega^3},
\end{equation}
where $\beta$, is a positive constant. According to
(\ref{d1.11.1}), this choice corresponds to a step function for
the susceptibility
\begin{equation}\label{d1.19.1}
\chi(t)=\left\{\begin{array}{ll}
\beta & t>0,\\
0 & t\leq 0,
\end{array}\right.
\end{equation}
and accordingly, the equation (\ref{d1.15}) can be written as
\begin{eqnarray}\label{d1.20}
&&m\ddot{q}+m\omega_0^2q\pm\beta\dot{q}=\xi(t),\nonumber\\
&&\xi(t)=\imath\sqrt{\frac{\beta\hbar c^3}{4\pi^2}}\int
\frac{d^3k}{\sqrt{ \omega_{\vec{k}}}} [b_{\vec{k}}(0)
e^{-i\omega_{\vec{k}}t}-b_{\vec{k}}^\dag(0)e^{i\omega_{\vec{k}}t}].\nonumber\\
&&
\end{eqnarray}
A complete solution of this equation for $t>0$ is as follows
\begin{eqnarray}\label{d1.21}
&&q(t)=e^{-\frac{\beta t}{2m}}
\{\frac{p(0)}{m\omega_1}\sin\omega_1 t +q(0)\cos\omega_1 t+
\frac{\beta}{2m\omega_1}q(0)\sin\omega_1 t\nonumber\\
&&-\frac{R(0)}{m\omega_1}\sin\omega_1 t -\frac{\beta
M(0)}{2m\omega_1}\sin\omega_1t-M(0)\cos\omega_1 t
-\frac{\dot{M}(0)}{\omega_1}\sin\omega_1 t\}+M(t),\nonumber\\
&&M(t)=\frac{\imath}{m}\sqrt{\frac{\beta\hbar c^3}{4 \pi^2}}\int
\frac{d^3k}{ \sqrt{\omega_{\vec{k}}}}[
\frac{b_{\vec{k}}(0)e^{-\imath \omega_{\vec{k}}t}
}{\omega_0^2-\omega_{\vec{k}}^2-\frac{\imath
\beta}{m}\omega_{\vec{k}}}+C.C],
\end{eqnarray}
where $\omega_1=\sqrt{\omega_0^2-\frac{\beta^2}{4m^2}}$. This
solution contains two parts, the  first part is an exponentially
decreasing function of time which is in fact the solution of the
homogeneous part of the equation (\ref{d1.20}), the second part is
a non-decaying term $M(t)$, which is the response to the noise
force $\xi(t)$. The part $M(t)$, does not have a classical
correspondence and is necessary for consistency of the
quantization of the dissipative systems.
 From (\ref{d1.6}),
(\ref{d1.10}), (\ref{d1.19.1}) and (\ref{d1.21}), we can obtain an
asymptotic answer for the conjugate momentum $p$ in the
large-time limit as
\begin{equation}\label{d1.22}
p(t)=m\dot{M}(t)+R_N(t)+\beta M(t)-\beta q(0).
\end{equation}
Let the state of the damped harmonic oscillator at $t=0$ be
$|\psi(0)\rangle=|0\rangle_B\otimes|n\rangle $ where $
|0\rangle_B $, is the vacuum state of the environment and $
|n\rangle $ is an excited state of the oscillator Hamiltonian,
i.e., $ H_s=\frac{p^2}{2m}+\frac{1}{2}m\omega_0^2q^2$, then it is
clear that
\begin{equation}\label{d1.23}
\langle\psi(0)| \frac{p^2(0)}{2m}+\frac{1}{2}m\omega_0^2 q^2(0)
|\psi(0)\rangle=(n+\frac{1}{2})\hbar\omega_0.
\end{equation}
  From (\ref{d1.21}) and (\ref{d1.22}), we find
\bea\label{d1.24}
 &&lim_{t\rightarrow\infty}[\langle\psi(0)| :
\frac{1}{2}m\dot{q}^2+\frac{1}{2}m\omega_0^2 q^2 :
|\psi(0)\rangle]\nonumber\\
&&=[\langle\psi(0)|: \frac{1}{2}m\dot{M}^2+\frac{1}{2}m\omega_0^2
M^2 : |\psi(0)\rangle] =0, \eea \\
where $:\hspace{0.4 cm}:$, is the normal ordering operator.
Therefore, macroscopically, analogous to classical mechanics, the
total energy of the particle tends to zero as time tends to
infinity. The expectation value of the environment Hamiltonian
($H_{B}$), in the same state ($|\psi(0)\rangle$), and large-time
limit, can be calculated by residue calculus as
 \bea\label{d1.25}
&&lim_{t\rightarrow\infty}[ \langle\psi(0)|
: H_B(t): |\psi(0)\rangle]\nonumber\\
 &&=\frac{\hbar\beta\omega_0}{\pi m}(n+\frac{1}{2})\int_0^\infty
\frac{\omega_{0}^2+x^2}{(\omega_0^2-x^2)^2+\frac{\beta^2}{m^2}x^2}dx=
(n+\frac{1}{2})\hbar\omega_{0}\nonumber.\\
&&. \eea\\
 So the total energy of the particle is absorbed by the environment
and which is a good signal of the consistency of the approach.
\subsection{Transition probabilities}
 The Hamiltonian (\ref{d1.1}), for a damped harmonic  oscillator can be written as
\begin{eqnarray}\label{d1.28}
&&H=H_0+H',\nonumber\\
&&H_0=(a^\dag a+\frac{1}{2})\hbar\omega_0+H_B, \nonumber\\
&&H'=-\frac{p}{m}R+\frac{R^2}{2m},
\end{eqnarray}
 where $a$ and $a^\dag$ are annihilation and creation operators of the Harmonic
 oscillator respectively.
 In interaction picture, we have
\bea\label{d1.29}
&& a_I(t)=e^{\frac{\imath}{\hbar}H_0 t}a(0)e^{-\frac{\imath}{\hbar}H_0 t}=a(0)
e^{-i\omega_0 t},\nonumber\\
&& b_{\vec{k}I}(t)=e^{\frac{\imath}{\hbar}H_0
t}b_{\vec{k}}(0)e^{-\frac{\imath}{\hbar}H_0
t}=b_{\vec{k}}(0)e^{-i\omega_{\vec{k}} t}, \eea\\
 the terms $\frac{R}{m} p $ and $\frac{R^2}{2m}$, are of the first and
second order of damping respectively, therefore, for a
sufficiently weak damping, $\frac{R^2}{2m}$ is small in
comparison with $\frac{R}{m}p$. Furthermore, in the first order
perturbation, $\frac{R^2}{2m}$, has not any role in those
transition probabilities where initial and final states of the
harmonic oscillator are different from each other, hence we can
neglect the term $\frac{R^2}{2m}$ in $H'$. Substituting $a_I$ and
$b_{\vec{k}I}$ from (\ref{d1.29}) in $ -\frac{ R}{m}p$ and using
the rotating wave approximation \cite{M5}, one can obtain $H_I'$
in interaction picture as
\begin{eqnarray}\label{d1.30}
&&H_I'=-\imath\sqrt{\frac{\hbar\omega_0}{2m}}\int_{-\infty}^{+\infty}d^3
k (f(\omega_{\vec{k}}) a^\dag b_{\vec{k}}(0)
e^{\imath(\omega_0-\omega_{\vec{k}})t}-f^*(\omega_{\vec{k}}) a
b_{\vec{k}}^\dag(0) e^{-\imath(\omega_0-\omega_{\vec{k}})t}.\nonumber\\
\end{eqnarray}
 In interaction picture, the density matrix $\rho(t)$, can be
 obtained from the initial density matrix $\rho(t_{0})$ as \cite{M6}
\begin{equation}\label{d1.31}
\rho_I(t)=U_I(t,t_0)\rho_I(t_0)U_I^\dag(t,t_0),
\end{equation}
where $U_I$, is the time evolution operator, which in the first
order perturbation is
\begin{eqnarray}\label{d1.32}
 U_I(t,t_0=0)&\approx & 1- \frac{\imath}{\hbar}\int_0^t d t_1 H'_I(t_1),\nonumber\\
 &=&1-\sqrt{\frac{\omega_0}{2m\hbar}}\int_{-\infty}^{+\infty}d^3 k
[f(\omega_{\vec{k}}) a^\dag b_{\vec{k}}(0)
e^{\frac{i(\omega_0-\omega_{\vec{k}})
t}{2}}\nonumber\\
&-&f^*(\omega_{\vec{k}}) a b_{\vec{k}}^\dag(0)
e^{\frac{-i(\omega_0-\omega_{\vec{k}})t}{2}}]
\frac{\sin\frac{(\omega_0-\omega_{\vec{k}})}{2}t}{\frac{(\omega_0-\omega_{\vec{k}})}{2}}.
\end{eqnarray}
Using the density operator $\rho_I$, we can calculate some
transition probabilities. for example let $
\rho_I(0)=|n\rangle\langle n|\otimes|0\rangle_{B\hspace{0.20
cm}B} \langle
 0|$, where $|0\rangle_B $, is the vacuum state of the reservoir and $ |n\rangle $,
 an exited state of the harmonic oscillator, then substituting $ U_I(t,0) $ from
 (\ref{d1.32}) in (\ref{d1.31}) and tracing out the environment degrees of freedom, we
 find
 \begin{eqnarray}\label{d1.33}
\rho_{sI}(t):&=&Tr_B(\rho_I(t)),\nonumber\\
 &=&|n\rangle\langle n|+\frac{n \omega_0}{2m\hbar}|n-1\rangle\langle
n-1|\int_{-\infty}^{+\infty} d^3 p |f(\omega_{\vec{p}})|^2
\frac{\sin^2\frac{(\omega_{\vec{p}}-\omega_0)}{2}t}
{(\frac{\omega_{\vec{p}}-\omega_0}{2})^2},\nonumber\\
&&
\end{eqnarray}
where we have used the formula $ Tr_B [|1_{\vec{k}}\rangle_B
\hspace{00.20cm}_B\langle1_{\vec{k'}}|
]=\delta(\vec{k}-\vec{k'})$. In large-time limit, we can write
$\frac{\sin^2\frac{(\omega_{\vec{p}}-\omega_0)}{2}t}
{(\frac{\omega_{\vec{p}}-\omega_0}{2})^2}\approx 2\pi
t\delta(\omega_{\vec{p}}-\omega_0)$, which leads to the following
relation for the density matrix
\begin{equation}\label{d1.34}
\rho_{sI}(t)=|n\rangle\langle n|+\frac{4\pi^2\omega_0^3 n
t|f(\omega_0)|^2}{m\hbar c^3}|n-1\rangle \langle n-1|.
\end{equation}
 From $ \rho_{sI}$ we can find The probability of transition
$|n\rangle\rightarrow|n-1\rangle$ as \\
\bea\label{d1.35}
&&\Gamma_{n\rightarrow n-1}=Tr [(|n-1\rangle \langle
n-1|)\rho(t)]\nonumber\\
&&=Tr_s[(|n-1\rangle_{\omega\hspace{0.20 cm}\omega}\langle
n-1|)\rho_{sI}(t)]=\frac{4\pi^2 \omega_0^3 n t
|f(\omega_0)|^2}{m\hbar c^3}, \eea\\
where $Tr_s $, denotes taking trace over harmonic oscillator
eigenstates. For the special choice (\ref{d1.19}), the transition
probability (\ref{d1.35}) becomes \\
\be\label{d1.36} \Gamma_{n\rightarrow n-1}=\frac{n\beta t}{m}.
\ee\\
In this case there is not any transition from the state $
|n\rangle $ to state $|n+1\rangle$.
 Now consider the case where the environment is an excited state in $t=0$, for
 example, take the density matrix as
\begin{equation}\label{d1.37}
\rho_I(0)=|n\rangle\langle
n|\otimes|1_{\vec{p}_1},...1_{\vec{p}_j}\rangle_{B}\hspace{0.20
cm} _{B} \langle
 1_{\vec{p}_1},...1_{\vec{p}_j}|,
\end{equation}
  where $|1_{\vec{p}_1},...1_{\vec{p}_j}\rangle_B $, denotes a state of
 the environment which contains $j$ quanta with the corresponding momenta
 $\vec{p}_1,...\vec{p}_j$,
 then using the relations
 \begin{eqnarray}\label{d1.38}
 &&Tr_B[ b_{\vec{k}}^\dag |1_{\vec{p}_1},...1_{\vec{p}_j}\rangle_{B\hspace{0.20 cm}B}\langle
 1_{\vec{p}_1},...1_{\vec{p}_j}| b_{\vec{k'}}]=\delta(\vec{k}-\vec{k'}),\nonumber\\
 &&Tr_B[b_{\vec{k}} |1_{\vec{p}_1},...1_{\vec{p}_j}\rangle_{B\hspace{0.20 cm}B} \langle
 1_{\vec{p}_1},...1_{\vec{p}_j}|b_{\vec{k'}}^\dag ]=
 \sum_{l=1}^j \delta(\vec{k}-\vec{p}_l)\delta(\vec{k'}-\vec{p}_l),
 \end{eqnarray}
we find in the large-time limit,
 \begin{eqnarray}\label{d1.39}
\rho_{sI}(t)&=&|n\rangle\langle
n|+\frac{(n+1)\omega_0}{2m\hbar}|n+1\rangle\langle n+1|
  \sum_{l=1}^j |f(\omega_{\vec{p}_l})|^2
\frac{\sin^2 \frac{(\omega_{\vec{p}_l}-\omega_0)}{2}t}{(
\frac{\omega_{\vec{p}_l}-\omega}{2})^2}\nonumber\\
&+&\frac{n\omega_0}{2m\hbar}|n-1\rangle \langle
n-1|\int_{-\infty}^{+\infty} d^3 k |f(\omega_{\vec{k}})|^2
\frac{\sin^2 \frac{(\omega_{\vec{k}}-\omega_0)}{2}t}{(
\frac{\omega_{\vec{k}}-\omega_0}{2})^2},
\end{eqnarray}
from which the transition probabilities $|n\rangle
\rightarrow|n-1\rangle $ and $|n\rangle\rightarrow |n+1\rangle $,
can be obtained as
\begin{eqnarray}\label{d1.40}
&&\Gamma_{n\rightarrow n-1}=Tr_s[|n-1\rangle \langle n-1|
\rho_{sI}(t)]=\frac{4\pi^2
\omega_0^3 n t}{m\hbar c^3}|f(\omega_0)|^2,\nonumber\\
&&\Gamma_{n\rightarrow n+1}=Tr_s[|n+1\rangle \langle n+1|
\rho_{sI}(t)]=\frac{(n+1)\pi t\omega_0}{m\hbar }|f(\omega_0)|^2
\sum_{l=1}^j \delta(\omega_{\vec{p}_l}-\omega_0).\nonumber\\
&&
\end{eqnarray}
Specially, for the choice (\ref{d1.19}), we have
\begin{eqnarray}\label{d1.41}
&&\Gamma_{n\rightarrow n-1}=\frac{n\beta
t}{m},\nonumber\\
&&\Gamma_{n\rightarrow n+1}=\frac{\beta(n+1)c^3 t}{4\pi m
\omega_0^2} \sum_{l=1}^j \delta( \omega_{\vec{p}_l}-\omega_0).
\end{eqnarray}
Another important case is when the environment has a canonical
thermal distribution, i.e.,
\begin{equation}\label{d1.42}
\rho_I(0)=|n\rangle\langle n| \otimes \rho_B^T, \hspace{2.00cm}
\rho_B^T=\frac{e^{\frac{-H_B}{K T}}}{TR_B(e^{\frac{-H_B}{KT}})},
\end{equation}
in this case, by applying the following relations
\begin{eqnarray}\label{d1.43}
&&Tr_B[ b_{\vec{k}}\rho_B^T b_{\vec{k'}}]=Tr_B[ b_{\vec{k}}^\dag
\rho_B^T
b_{\vec{k'}}^\dag]=0,\nonumber\\
&& Tr_b[b_{\vec{k}}\rho_B^T
b_{\vec{k'}}^\dag]=\frac{\delta(\vec{k}-\vec{k'})}
{e^{\frac{\omega_{\vec{k}}}{K T}}-1},\nonumber\\
&&Tr_B[ b_{\vec{k}}^\dag \rho_B^T
b_{\vec{k'}}]=\frac{\delta(\vec{k}-\vec{k'})e^{\frac{\omega_{\vec{k}}}{K
T}}}{e^{\frac{\omega_{\vec{k}}}{K T}}-1},
\end{eqnarray}
we find the density operator $\rho_{sI}(t)$ in interaction
picture as
 \begin{eqnarray}\label{d1.44}
 &&\rho_{sI}(t):=Tr_B [\rho_I(t)]\nonumber\\
 &&=|n\rangle\langle n|+\frac{(n+1)\omega_0}{2m\hbar}|n+1\rangle \langle n+1|
\int_{-\infty}^{+\infty} d^3 k
\frac{|f(\omega_{\vec{k}})|^2}{e^{\frac{\omega_{\vec{k}}}{K
T}}-1} \frac{\sin^2 \frac{(\omega_{\vec{k}}-\omega_0)}{2}t}{
(\frac{\omega_{\vec{k}}-\omega_0}{2})^2}\nonumber\\
&&+\frac{n\omega_0}{2m\hbar}|n-1\rangle \langle
n-1|\int_{-\infty}^{+\infty} d^3 k
\frac{|f(\omega_{\vec{k}})|^2e^{\frac{\omega_{\vec{k}}}{K
T}}}{e^{\frac{\omega_{\vec{k}}}{K T}}-1} \frac{\sin^2
\frac{(\omega_{\vec{k}}-\omega_0)}{2}t}{(
\frac{\omega_{\vec{k}}-\omega_0}{2})^2},
\end{eqnarray}
which accordingly lead to the following transition probabilities
in large-time limit
\begin{eqnarray}\label{d1.45}
&&\Gamma_{n\rightarrow n-1}=Tr_s[|n-1\rangle \langle n-1|
\rho_{sI}(t)]=\frac{4\pi^2\omega_0^3 n t}{m\hbar c^3
}\frac{|f(\omega_0)|^2 e^{\frac{\omega_0}{K
T}}}{e^{\frac{\omega_0}{K T}}-1},\nonumber\\
&&\Gamma_{n\rightarrow n+1}=Tr_s[|n+1\rangle\langle n+1|
\rho_{sI}(t)]=\frac{4\pi^2\omega_0^3 (n+1)
t}{m\hbar c^3}\frac{|f(\omega_0)|^2}{e^{\frac{\omega_0}{K T}}-1},\nonumber\\
&&
\end{eqnarray}
and for the special case (\ref{d1.19}), are reduced to
\begin{eqnarray}\label{d36}
&&\Gamma_{n\rightarrow n-1}=\frac{n\beta te^{\frac{\omega_0}{K
T}}}{m(e^{\frac{\omega_0}{K T}}-1)},\nonumber\\
&&\Gamma_{n\rightarrow n+1}=\frac{(n+1)\beta
t}{m(e^{\frac{\omega_0}{K T}}-1)}.
\end{eqnarray}
 Therefore, in low temperatures regime, the energy flows from the oscillator to the environment
 by the rate $\Gamma_{n\rightarrow n-1}\mapsto\frac{n\beta}{m}$, and no energy flows from
 the environment to the oscillator in this case.
\section{Three-dimensional quantum dissipative systems}
In this section, we study a general three-dimensional quantum
dissipative system. This time for generality, we model the
environment of the system by two independent quantum fields,
namely $B$ and $\tilde{B}$ quantum fields. Quantum field $B$,
interacts with the momentum of the main system through a minimal
coupling term and the quantum field $\tilde{B}$, interacts with
the position operator of the main system similar to a dipole
interaction term. For this purpose  we take the Hamiltonian of
the environment as
\begin{eqnarray}\label{d2.1}
&&H_E= H_B(t)+H_{ \tilde{B}}(t),\nonumber\\
&&H_B(t)=\int_{-\infty}^{+\infty}d^3k \hbar\omega_{\vec{k}}
b_{\vec{k}}^\dag(t) b_{\vec{k}}(t),\nonumber\\
&&H_{ \tilde{B}}(t)=\int_{-\infty}^{+\infty}d^3k
\hbar\omega_{\vec{k}} d_{\vec{k}}^\dag(t) d_{\vec{k}}(t),
\end{eqnarray}
where according to the explanations of the previous section, we
choose   $\omega_{\vec{k}}=c|\vec{k}|$. The annihilation and
creation operators $b_{\vec{k}}$, $b_{\vec{k}}^\dag$,
$d_{\vec{k}}$ and $d_{\vec{k}}^\dag$, satisfy the commutation
relations
\begin{eqnarray}\label{d2.2}
&&[b_{\vec{k}}(t),b_{\vec{k}'}^\dag(t)]=\delta(\vec{k}-\vec{k}'),\nonumber\\
&&[d_{\vec{k}}(t),d_{\vec{k}'}^\dag(t)]=\delta(\vec{k}-\vec{k}'),
\end{eqnarray}
 and the rest of commutation relations are zero.

  Let the damped system be a
particle with mass $m$ under an external potential $v(\vec{x})$,
we take the total Hamiltonian, i.e., system plus the environment,
as
\begin{equation} \label{d2.3}
H=\frac{(\vec{p}-\vec{R})^2}{2m}+v(\vec{x})-\vec{\tilde{R}}\cdot\vec{x}+H_B
+H_{ \tilde{B}},
\end{equation}
where $\vec{x}$ and $\vec{p}$, are position and canonical
conjugate momentum operators of the particle respectively, and
satisfy the canonical commutation rules
\begin{equation}\label{d2.4}
[x_i,p_j]=\imath\hbar\delta_{ij}.
\end{equation}
Operators $\vec{R}$ and $\vec{\tilde{R}}$, play the basic role in
interaction between the system and the environment and are defined
by
\begin{eqnarray}\label{d2.5}
&&\vec{R}(t)=\int_{-\infty}^{+\infty}d^3k [f(\omega_{\vec{k}})
b_{\vec{k}}(t)+f^*(\omega_{\vec{k}})b_{\vec{k}}^\dag(t)]\hat{k},\nonumber\\
&&\vec{\tilde{R}}(t)=\int_{-\infty}^{+\infty}d^3k
[g(\omega_{\vec{k}})
d_{\vec{k}}(t)+g^*(\omega_{\vec{k}})d_{\vec{k}}^\dag(t)]\hat{k},
\end{eqnarray}
where $\hat{k}=\frac{\vec{k}}{|\vec{k}|}$. For future purposes the
fields $\vec{R}$ and $\vec{\tilde{R}}$, are taken to have the
longitudinal polarization. et us call the functions
$f(\omega_{\vec{k}})$ and $g(\omega_{\vec{k}})$, the coupling
functions between the environment and the system. The Heisenberg
equations for the position $\vec{x}(t)$ and the momentum
$\vec{p}$ of the particle are
\begin{eqnarray}\label{d2.6}
&&\dot{\vec{x}}=\frac{\imath}{\hbar}[H,\vec{x}]=\frac{\vec{p}-\vec{R}}{m},\nonumber\\
&&
\dot{\vec{p}}=\frac{\imath}{\hbar}[H,\vec{p}]=-\vec{\nabla}v+\vec{\tilde{R}}.
\end{eqnarray}
By eliminating $\vec{p}$ in relations (\ref{d2.6}), we come to the
following equation for the damped system
\begin{equation}\label{d2.7}
m\ddot{\vec{x}}=-\vec{\nabla}v-\dot{\vec{R}}+\vec{\tilde{R}}.
\end{equation}
 Using (\ref{d2.2}), one can easily find the Heisenberg equations for the operators
 $ b_{\vec{k}}$ and $d_{\vec{k}}$ as
\begin{eqnarray}\label{d2.8}
&&\dot{b}_{\vec{k}}=\frac{\imath}{\hbar}[H,b_{\vec{k}}]=-\imath\omega_{\vec{k}}
b_{\vec{k}}+\frac{\imath}{\hbar}f^*(\omega_{\vec{k}})\hat{k}\cdot\dot{\vec{x}},\nonumber\\
&&\dot{d}_{\vec{k}}=\frac{\imath}{\hbar}[H,d_{\vec{k}}]=-\imath\omega_{\vec{k}}
d_{\vec{k}}+\frac{\imath}{\hbar}g^*(\omega_{\vec{k}})\hat{k}\cdot\vec{x}.
\end{eqnarray}
These equations can be solved formally as
\begin{eqnarray}\label{d2.9}
&&b_{\vec{k}}(t)=b_{\vec{k}}(0)e^{-\imath\omega_{\vec{k}}
t}+\frac{\imath}{\hbar}f^*(\omega_{\vec{k}})\hat{k}\cdot\int_0^t
d t'
e^{-\imath\omega_{\vec{k}}(t-t')} \dot{\vec{x}}(t'),\nonumber\\
&&d_{\vec{k}}(t)=d_{\vec{k}}(0)e^{-\imath\omega_{\vec{k}}
t}+\frac{\imath}{\hbar}g^*(\omega_{\vec{k}})\hat{k}\cdot\int_0^t
d t'e^{-\imath\omega_{\vec{k}}(t-t')} \vec{x}(t').\nonumber\\
\end{eqnarray}
Substituting these solutions into (\ref{d2.5}), one obtains what
we have called the constitutive equations of the environment,
i.e.,
\begin{eqnarray}\label{d2.10}
&&\vec{R}(t)=\vec{R}_N(t)+\int_0^{|t|}
dt'\chi(|t|-t')\dot{\vec{x}}(\pm t'),\nonumber\\
&&\vec{\tilde{R}}(t)=\vec{\tilde{R}}_N(t)+\int_0^{|t|}
dt'\tilde{\chi}(|t|-t')\vec{x}(\pm t'),
\end{eqnarray}
where the upper(lower) sign, corresponds to $t>0$$(t<0)$,
respectively. The functions
\begin{eqnarray}\label{d2.11}
&&\chi(t)=\frac{16\pi}{3\hbar c^3}\int_0^\infty d \omega
\omega^2|f(\omega)|^2\sin\omega t \hspace{1.50cm}t>0,\nonumber\\
&&\chi(t)=0\hspace{6.2500cm}t\leq0,
\end{eqnarray}
and
\begin{eqnarray}\label{d2.12}
&&\tilde{\chi}(t)=\frac{16\pi}{3\hbar c^3}\int_0^\infty d \omega
\omega^2|g(\omega)|^2\sin\omega t \hspace{1.50cm}t>0,\nonumber\\
&&\tilde{\chi}(t)=0\hspace{6.2500cm}t\leq0,
\end{eqnarray}
are called the susceptibilities of the environment. The operators
 $\vec{R}_N$ and $\vec{\tilde{R}}_N$, are the noise operators
\begin{equation}\label{d2.13}
\vec{R}_N(t)=\int_{-\infty}^{+\infty} d^3 k
[f(\omega_{\vec{k}})b_{\vec{k}}(0)
e^{-\imath\omega_{\vec{k}}t}+f^*(\omega_{\vec{k}})
b_{\vec{k}}^\dag(0)e^{\imath\omega_{\vec{k}}t}]\hat{k},
\end{equation}
\begin{equation}\label{d2.14}
\vec{\tilde{R}}_N(t)=\int_{-\infty}^{+\infty} d^3 k
[g(\omega_{\vec{k}})d_{\vec{k}}(0)
e^{-\imath\omega_{\vec{k}}t}+g^*(\omega_{\vec{k}})d_{\vec{k}}^\dag(0)
e^{\imath\omega_{\vec{k}}t}]\hat{k}.
\end{equation}
If we are given the macroscopic susceptibilities $\chi(t)$ and $
\tilde{\chi}(t)$, which are zero for $t\leq 0 $, then we can
invert the equations (\ref{d2.11}) and (\ref{d2.12}) and obtain
the corresponding coupling functions $ f(\omega) $ and $g(\omega)$
as
\begin{eqnarray}\label{d2.15}
|f(\omega)|^2&=&\frac{3\hbar c^3 }{8\pi^2\omega^2}\int_0^\infty
dt\chi(t)
\sin\omega t,\hspace{1cm}\omega>0,\nonumber\\
|f(\omega)|^2&=&0,\hspace{1cm}\omega=0.
\end{eqnarray}
\begin{eqnarray}\label{d2.16}
|g(\omega)|^2&=&\frac{3\hbar c^3 }{8\pi^2\omega^2}\int_0^\infty
dt\tilde{\chi}(t)
\sin\omega t,\hspace{1cm}\omega>0,\nonumber\\
|g(\omega)|^2&=&0,\hspace{1cm}\omega=0.
\end{eqnarray}
It should be noted that the sign of the right hand side of the
formulas (\ref{d2.15}) and (\ref{d2.16}) should be positive, this
is because, the right hand side of these formulas is proportional
to the imaginary part of the susceptibilities $ \chi(t) $ and
$\tilde{\chi}(t)$ in the frequency domain and this imaginary part
is also connected to energy losses of a system in the presence of
an environment, so it is necessarily positive. If these integrals
would not be positive for all frequencies, the susceptibilities
$\chi(t)$ and $\tilde{\chi}(t)$ should be discarded, as they
would be unphysical. Substituting $\vec{R}$ and
$\vec{\tilde{R}}$, from the constitutive equations (\ref{d2.10}),
into (\ref{d2.7}), one can obtain the equation of motion of the
operator $\vec{x}$ as \\
\be\label{d2.17}
 m\ddot{\vec{x}}\pm\frac{d}{d t}\int_0^{|t|} d
t'\chi(|t|-t')\dot{\vec{x}}(\pm t')-\int_0^{|t|} d
t'\tilde{\chi}(|t|-t')\vec{x}(\pm t')=-\vec{\nabla}
v(\vec{x})-\dot{\vec{R}}_N+\vec{\tilde{R}}_N, \ee\\
where the upper (lower) sign corresponds to $ t>0 $ $(t<0)$. This
is the generalized macroscopic Langevin-Schr\"{o}dinger equation
for a dissipative quantum system. Therefore, coupling with the
environment in microscopic level, is described macroscopically
 by two random forces $-\dot{\vec{R}}_N$ and $\vec{\tilde{R}}_N $ and the
 corresponding memory functions $\chi$ and $\tilde{\chi}$.

  As a simple example, let us consider a damping three-dimensional
harmonic oscillator with resonance frequency $\omega_0$ and mass
$m$, where the damping force is proportional to the velocity and
position of the oscillator. In this case,
 we take the susceptibilities $ \chi(t)$ and $\tilde{\chi}(t)$, as
 follows
\begin{displaymath}
\chi(t)=\left\{\begin{array}{ll}
\beta & t>0,\\
0 & t\leq 0,
\end{array}\right.
\end{displaymath}
\begin{displaymath}
\tilde{\chi}(t)=\left\{\begin{array}{ll}
\frac{\alpha m \omega_0^2}{\triangle} & 0<t<\triangle,\\
0 & \textrm{otherwise},
\end{array}\right.
\end{displaymath}
where $ 0<\alpha<1$ and $\beta$ are some positive constants. Now
one can use the relations (\ref{d2.15}) and (\ref{d2.16}) to
obtain the related coupling functions as
\begin{equation}\label{d2.18}
|f(\omega)|^2=\frac{3\hbar c^3\beta}{8\pi^2\omega^3},\hspace{2.00
cm}\omega\neq0,
\end{equation}
\begin{equation}\label{d2.19}
|g(\omega)|^2=\frac{3\hbar c^3\alpha m\omega_0^2}{8\pi^2
\omega^2}\frac{\sin^2\frac{\omega\triangle}{2}}
{\frac{\omega\triangle}{2}},\hspace{1.00cm}\omega\neq 0.
\end{equation}
In this case, the Langevin-Schr\"{o}dinger equation (\ref{d2.17}),
is reduced to
\begin{eqnarray}\label{d2.20}
 m\ddot{\vec{x}}+m\omega_0^2\vec{x}\pm \beta\dot{\vec{x}}
-\frac{\alpha m\omega_0^2}{\triangle}\int_{|t|-\triangle}^{|t|} d
t'\vec{x}(\pm t')=-\dot{R}_N+\vec{\tilde{R}}_N,\nonumber\\
 &&
\end{eqnarray}
where $ \vec{R}_N$ and $\vec{\tilde{R}}$ are the noise operators
(\ref{d2.13}) and (\ref{d2.14}) with coupling functions
(\ref{d2.18}) and (\ref{d2.19}) respectively. It is clear that in
the limit $\triangle\rightarrow 0$, the coupling function
(\ref{d2.19}), tends to zero and accordingly, the noise force
$\vec{\tilde{R}}_N$ vanishes. Therefore, the equation
(\ref{d2.20}) becomes
\begin{eqnarray}\label{d2.21}
 &&m\ddot{\vec{x}}+(1-\alpha)m\omega_0^2\vec{x}\pm
 \beta\dot{\vec{x}}=\vec{\xi}(t),\nonumber\\
 &&\vec{\xi}(t)=\imath\sqrt{\frac{3\beta\hbar c^3}{8\pi^2}}\int
d^3k[\int_{-\infty}^{+\infty}\frac{ d^3 k}{\sqrt{
\omega_{\vec{k}}}} [f(\omega_{\vec{k}})b_{\vec{k}}(0)
e^{-i\omega_{\vec{k}}t}-f^*(\omega_{\vec{k}})b_{\vec{k}}^\dag(0)
e^{i\omega_{\vec{k}}t}].\nonumber\\
&&
\end{eqnarray}
For an absorptive environment, $0<\alpha<1$, otherwise the
environment would behave as an amplifier, i.e., energy transfers
to the system from the environment, this may be an indication for
making other related models, but in our discussion, this case is
not physical. As is seen from this equation, the environment
affect the motion of the particle by two damping forces, one is a
friction force proportional to the velocity and the other is a
force proportional to the position and in opposite direction of
the external force $-\vec{\nabla}v(\vec{x})=-m\omega_0^2\vec{x}$.
In fact, the later damping force, influence the frequency of the
oscillator.
\subsection{Dissipative quantum two-level systems}
The rate of spontaneous decay of a two-level quantum system
imbedded in an absorptive environment, can be calculated in this
approach as follows. Here we assume that the interaction between
the particle and its environment is so weak that we can use the
Weiskopf-Wigner approximation \cite{M5} for calculating the decay
constant. Let us write the total Hamiltonian (\ref{d2.3}), as the
following form
\begin{eqnarray}\label{d2.22}
&&H=H_0+H',\nonumber\\
&&H_0=\frac{\vec{p}^2}{2m}+v(
\vec{x})+H_B+H_{\tilde{B}},\nonumber\\
&&H'=-\frac{\vec{p}}{m}\cdot\vec{R}-\vec{\tilde{R}}\cdot\vec{x},
\end{eqnarray}
where we have ignored from the term $ \frac{\vec{R}^2}{2m} $,
because, this term does not contribute in the following
calculations of the
 decaying rate of the system. Now for a considerable simplification, we
 restrict ourselves to a two-state model of a system. In fact,
the Hilbert space of the main system is artificially truncated to
the two states $|1\rangle$ and $|2\rangle$ with unperturbed
energy eigenvalues $ E_1$, $E_2 $, respectively. Therefore, for
the Hamiltonian of the system we can write
\begin{equation}\label{d2.23}
H_{s}=
\frac{\vec{p}^2}{2m}+v(\vec{x})\equiv\frac{1}{2}\hbar\Omega_0\sigma_z,
\hspace{1.00 cm} \sigma_z= |2\rangle\langle
2|-|1\rangle\langle1|,\hspace{1.00cm}\Omega_0=\frac{E_2-E_1}{\hbar},
\end{equation}
and the interaction Hamiltonian $ H' $ becomes
\begin{eqnarray}\label{d2.24}
H'&=&-\imath\Omega_0\int d^3k[\sigma b_{\vec{k}}f(
\omega_{\vec{k}})\vec{x}_{12}\cdot\hat{k}+\sigma b_{\vec{k}}^\dag
f^*(\omega_{\vec{k}})\vec{x}_{12}\cdot\hat{k}-C.C]\nonumber\\
&-&\int d^3k[\sigma d_{\vec{k}}g(
\omega_{\vec{k}})\vec{x}_{12}\cdot\hat{k}+\sigma d_{\vec{k}}^\dag
g^*( \omega_{\vec{k}})\vec{x}_{12}\cdot\hat{k}+C.C],
\end{eqnarray}
where
\begin{equation}\label{d2.25}
 \sigma=|1\rangle\langle 2|, \hspace{2.00 cm}\vec{x}_{12}=\langle
1|\vec{x}|2\rangle,
\end{equation}
and we have used this fact that the energy eigenstates of the
Hamiltonian (\ref{d2.23}) have a well defined parity, so that the
diagonal elements of the operators $\vec{x}$ and $\vec{p}$, are
zero.

 To study the spontaneous decay of an initially excited
two-level system, imbedded in an absorptive environment, we may
look for the system wave function at time $t$ in interaction
picture
\begin{eqnarray}\label{d2.26}
&& |\psi(t)\rangle_I=c(t)|2\rangle|0\rangle_B|0\rangle_{
\tilde{B}} +
\int d^3k D_{\vec{k}}(t) |1\rangle|\vec{k}\rangle_B|0\rangle_{\tilde{B}}+\nonumber\\
&&\int d^3\vec{k}
G_{\vec{k}}(t)|1\rangle|0\rangle_B|\vec{k}\rangle_{ \tilde{B}},
\end{eqnarray}
where $ |0\rangle_B$ and $|0\rangle_{\tilde{B}}$, are vacuum
states of quantum fields $ B$ and $\tilde{B}$, respectively, and
the coefficient $c(t)$, $ D_{\vec{k}}(t)$ and $G_{\vec{k}}(t)$,
are to be specified by the Schr\"{o}dinger equation
\begin{equation}\label{d2.27}
\imath\hbar\frac{\partial |\psi(t)\rangle_I}{\partial
t}=H'_I(t)|\psi(t)\rangle_I,
\hspace{1.00cm}H'_I(t)=e^{\frac{\imath
H_0^st}{\hbar}}H'^se^{\frac{-\imath H_0^st}{\hbar}},
\end{equation}
with initial conditions $c(0)=1$, $
D_{\vec{k}}(0)=G_{\vec{k}}(0)=0$. Where $ H_0^s $ and $ H'^s $ are
Hamiltonians $H_0$ and $H'$ in the Schr\"{o}dinger picture
respectively. Substituting $|\psi(t)\rangle_I$ from
(\ref{d2.26}), in (\ref{d2.27}), and using the rotating wave
approximation \cite{M6}, we find the differential equations
\begin{eqnarray}\label{d2.28}
\imath\hbar\dot{c}(t)&=&-\int
d^3k[-\imath\Omega_0\vec{x}^*_{12}\cdot\hat{k}f(
\omega_{\vec{k}})e^{\imath(\Omega_0-\omega_{\vec{k}})t}D_{\vec{k}}(t)\nonumber\\
&+&\vec{x}^*_{12}\cdot \hat{k}g( \omega_{\vec{k}})
e^{i(\Omega_0-\omega_{\vec{k}})t}G_{\vec{k}}(t)],
\end{eqnarray}
\begin{equation}\label{d2.29}
\imath\hbar\dot{D}_{\vec{k}}(t)=-\imath\Omega_0\vec{x}_{12}\cdot\hat{k}f^*(
\omega_{\vec{k}}) e^{-\imath(\Omega_0-\omega_{\vec{k}})t}c(t),
\end{equation}
\begin{equation}\label{d2.30}
\imath\hbar\dot{G}_{\vec{k}}(t)=-\vec{x}_{12}\cdot\hat{k}g^*(
\omega_{\vec{k}}) e^{-\imath(\Omega_0-\omega_{\vec{k}})t}c(t),
\end{equation}
One can solve this differential equations using the Laplace
transformation technique and deduce
\begin{eqnarray}\label{d2.31}
\dot{c}(t)&=&\int_0^t d t'\gamma(t-t')c(t'),\nonumber\\
\gamma(t-t') &=& -\int d^3k[ \Omega_0^2|\vec{x}_{12}\cdot\hat{k}f(
\omega_{\vec{k}})|^2+|\vec{x}_{12}\cdot\hat{k}g(
\omega_{\vec{k}})|^2]e^{\imath(
\Omega_0-\omega_{\vec{k}})(t-t')}.\nonumber\\
&&
\end{eqnarray}
Here we restrict our attention to the weak coupling regime, where
the Markov approximation \cite{M7} applies. That is to say, we may
replace $c(t')$ in the integrand of (\ref{d2.31}) with $c(t)$ and
approximate the time integral $\int_0^t dt'\gamma(t-t')$, in
large-time limit, as
\begin{equation}\label{d2.32}
\int_0^t dt' e^{i(\Omega_0-\omega_{\vec{k}})(t-t')}\approx
[iP\frac{1}{\Omega_0-\omega_{\vec{k}}}+
\pi\delta(\Omega_0-\omega_{\vec{k}})],
\end{equation}
where $P$, denotes th principal Cauchy value. After some algebraic
calculations, we find
\begin{equation}\label{d2.33}
\dot{c}(t)=-(\beta+i\Delta)c(t),
\end{equation}
where $\beta$ is the decay constant
\begin{eqnarray}\label{d2.34}
&&\beta =\frac{4\pi^2
\Omega_0^4|\vec{x}_{12}|^2|f(\Omega_0)|^2}{3\hbar^2
c^3}+\frac{4\pi^2 |\vec{x}_{12}|^2|g(\Omega_0)|^2}{3\hbar^2
c^3},\nonumber\\
&&
\end{eqnarray}
and $ \Delta $ is the shift frequency
\begin{eqnarray}\label{d2.35}
&&\Delta
=\frac{4\pi|\vec{x_{12}}|^2}{3\hbar^{2}c^3}P\int_{0}^{\infty}d\omega
\frac{[\Omega_{0}^2|f(\omega)|^2
+|g(\omega)|^2]\omega^2}{\Omega_{0}-\omega}.
\end{eqnarray}
\section{Dissipative scalar field theory}
In this section, we discuss the dissipative scalar field theory.
We take for convenience a 1+1-dimensional scalar field but the
generalization to a general scalar field theory is
straightforward. For this purpose, assume $\psi(x,t)$ be a scalar
field operator defined on a closed compact interval $[0,L]$,
satisfying the boundary conditions $ \psi(0)=\psi(L)=0$. The
field $\psi(x,t)$, can be expanded in terms of the orthogonal
wave function $\sin\frac{n\pi x }{L}$, as
\begin{equation} \label{d3.1}
\psi(x,t)=\sum_{n=1}^\infty\sqrt{\frac{\hbar}{L\lambda\omega_n}}
[a_n(t)+a_n^\dag(t)]\sin\frac{n\pi
x}{L},
\end{equation}
where $\omega_n=\sqrt{\frac{\mu}{\lambda}}\frac{n\pi}{L}$, and
 $\lambda,\mu $, are some constants. The operators $a_{n}$ and $ a_n^\dag $,
  are annihilation and creation operators of the scalar field $\psi(x,t)$,
  respectively and satisfy the following equal-time commutation
  relations
\begin{equation}\label{d3.2}
[a_n(t),a_m^\dag(t)]=\delta_{nm}.
\end{equation}
The conjugate canonical momentum density of the field $\psi(x,t)$,
can be also expanded in the same basis as
\begin{equation}\label{d3.3}
\pi_\psi(x,t)=i\sum_{n=1}^\infty\sqrt{\frac{\hbar\lambda\omega_n}{L}}
[a_n^\dag(t)-a_n(t)]\sin\frac{n\pi x}{L}.
\end{equation}
 From (\ref{d3.2}), we deduce that $\psi$, $\pi_\psi$, satisfy the equal time
commutation relation
\begin{equation}\label{d3.4}
[\psi(x,t),\pi_\psi(x',t)]=i\hbar\delta(x-x').
\end{equation}
The Hamiltonian of the scalar field $\psi(x,t)$, in normal ordered
form is
\begin{equation}\label{d3.5}
H_s=\int_0^L d x:[\frac{\pi_\psi^2}{2\lambda
}+\frac{1}{2}\mu\psi_x^2]:=\sum_{n=1}^\infty\hbar\omega_n
a_n^\dag(t)a_n(t),
\end{equation}
 where $\psi_x$, denotes the derivative with respect to $x$.
Now similar to what we did in previous sections, here we model
the environment
 by a quantum field $B$ containing an infinite number, but numerable, of Klein-Gordon
 fields. Therefore, the environment Hamiltonian, can be written as
\begin{equation}\label{d3.6}
H_B(t)=\sum_{n=1}^\infty\int_{-\infty}^{+\infty}d^3k\,
\hbar\omega_{\vec{k}}\,b_{n\vec{k}}^\dag(t) b_{n\vec{k}}(t),
\hspace{1.50 cm} \omega_{\vec{k}}=c|\vec{k}|,
\end{equation}
where $b_{n\vec{k}}$ and $b_{n\vec{k}}^\dag$, are annihilation and
creation operators of the environment respectively and satisfy
the following bosonic commutation relations (bosonic bath)
\begin{equation}\label{d3.7}
[b_{n\vec{k}}(t),b_{m\vec{k}'}^\dag(t)]=\delta_{nm}\delta(\vec{k}-\vec{k}'),
\end{equation}
In a systematic way, and also according to what we did in the
previous sections, we write the total Hamiltonian, i.e., the
scalar field $\psi(x,t)$ plus the environment, as
\begin{equation}\label{d3.8}
H(t)=\int_0^L d x
\frac{[\pi_\psi(x,t)-R(x,t)]^2}{2\lambda}+\frac{1}{2}\mu\psi_x^2+H_B,
\end{equation}
wherein the operator $R(x,t)$, has the basic role in interaction
between the scalar field and its environment and is defined by
\begin{equation}\label{d3.9}
R(x,t)=\sqrt{\frac{2}{L}}\sum_{n=1}^\infty\int_{-\infty}^{+\infty}d^3k
[f(\omega_{\vec{k}},x)
b_{n\vec{k}}(t)+f^*(\omega_{\vec{k}},x)b_{n\vec{k}}^\dag(t)]\sin\frac{n\pi
x }{L}.
\end{equation}
 Let us call the function $f(\omega_k,x)$, the coupling function between the scalar field
 and its environment. The coupling function is independent(dependent) on $x$ for a
 homogeneous(inhomogeneous) environment respectively.

  One can easily show that the Heisenberg equations for $\psi(x,t)$
 and $ \pi_\psi(x,t)$ are
\begin{eqnarray}\label{d3.91}
&&\dot{\psi}(x,t)=\frac{\imath}{\hbar}[H,\psi]=\frac{\psi(x,t)-R(x,t)}{\lambda},\nonumber\\
&&\dot{\pi}_\psi(x,t)=\frac{\imath}{\hbar}[H,\pi_\psi]=-\mu\psi_{xx},
\end{eqnarray}
we can eliminate $\pi_\psi$ between the relations (\ref{d3.91})
and finally find the following equation for the field $\psi$
\begin{equation}\label{d3.92}
\lambda\ddot{\psi}-\mu\psi_{xx}=-\dot{R}.
\end{equation}
 Now similar to the previous sections, if we write the Heisenberg equation for the annihilation
 operator $b_{n\vec{k}}$ and solve it formally then substitute this solution into
 (\ref{d3.9}), we can obtain a constitutive equation for the environment as
\begin{equation}\label{d3.10}
R(x,t)=R_N(x,t)+\int_0^{|t|}dt'\chi(x,|t|-t')\dot{\psi}(x,\pm t'),
\end{equation}
where the upper(lower) sign, corresponds to $ t>0$($t<0$)
respectively and
\begin{eqnarray}\label{d3.11}
&&\chi(x,t)=\frac{8\pi}{\hbar c^3}\int_0^\infty d \omega
\omega^2|f(\omega,x)|^2\sin\omega t, \hspace{1.50cm}t>0,\nonumber\\
&&\chi(x,t)=0,\hspace{6.8cm}t\leq 0,
\end{eqnarray}
is the susceptibility of the environment. By inverting the
relation (\ref{d3.11}), we can obtain the coupling function
$f(\omega ,x)$, in terms of the susceptibility, as
\begin{eqnarray}\label{d1.12}
|f(\omega,x)|^2&=&\frac{\hbar c^3 }{4\pi^2\omega^2}\int_0^\infty
dt\chi(x,t)
\sin\omega t,\hspace{1cm}\omega>0,\nonumber\\
|f(\omega,x)|^2&=&0,\hspace{5.35cm}\omega=0.
\end{eqnarray}
The operator
\begin{equation}\label{d3.13}
R_N(x,t)=\sqrt{\frac{2}{L}}\sum_{n=1}^\infty\int_{-\infty}^{+\infty}d^3k
[f(\omega_{\vec{k}},x)
b_{n\vec{k}}(0)e^{-\imath\omega_{\vec{k}}t}+f^*(\omega_{\vec{k}},x)b_{n\vec{k}}^\dag(0)e^{\imath\omega_{\vec{k}}t}]\sin\frac{n\pi
x }{L},
\end{equation}
is a noise operator and necessary for a consistent quantization
of a dissipative theory.
 Finally, substituting (\ref{d3.10}) in (\ref{d3.92}), leads to
 the following wave equation for the field operator $\psi $
\begin{equation}\label{d3.14}
\lambda\ddot{\psi}-\mu\psi_{xx}\pm\frac{d}{dt}\int_0^{|t|}d
t'\chi(x,|t|-t')\dot{\psi}(x,\pm t')=-\dot{R}_N(x,t).
\end{equation}
Taking the Laplace transformation of this equation, we find that
the forward and backward Laplace transforms of $\psi(x,t)$,
satisfy the following equation

\be\label{d3.15} [\lambda
s^2+s^2\underline{\chi}(x,s)]\underline{\psi}^{f,b}(x,s)-\mu\underline{\psi}^
{f,b}_{xx}(x,s)=\underline{J}^{f,b}(x,s)
\ee
where

$$\underline{J}^{f,b}(x,s)=[\lambda s
+s\underline{\chi}(x,s)]\psi(x,0)\mp \lambda
\dot{\psi}(x,0)\mp\underline{R}_N^{f,b}(x,s)\pm\dot{R}_N(x,0),$$

 the upper(lower) sign, corresponds to the forward(backward)
Laplace transformations respectively and $\underline{\chi}(x,s)$,
is the Laplace transform of the susceptibility. One can solve
equation (\ref{d3.15}) in terms of the Green function
\begin{equation}\label{d3.16}
 \underline{\psi}^{f,b}x,s)=\int_0^L dx'G(x,x',s)\underline{J}^{f,b}(x',s),
 \end{equation}
 where $ G(x,x',s)$ satisfies
\begin{equation}\label{d3.17}
(\lambda s^2+s^2\underline{\chi}(s,x)) G(x,x',s) -\mu
G_{xx}(x,x',s)=\delta(x-x'),
\end{equation}
with the boundary conditions $ G(0,x',s)=G(L,x',s)=0$.  For a
homogeneous environment, the coupling function and the
susceptibility $\chi$, are position independent and in this case,
the Green function is
\begin{equation}\label{d3.18}
G(x,x',s)=\frac{2}{L}\sum_{n=1}^\infty\frac{1}{\lambda
s^2+s^2\underline{\chi}(s)+\lambda \omega_n^2}\sin\frac{n\pi
x}{L}\sin\frac{n \pi x'}{L}.
\end{equation}
\subsection{A special case}
For the special choice of coupling function (\ref{d1.19}),
equation (\ref{d3.14}), becomes
\begin{eqnarray}\label{d3.19}
&&\lambda\ddot{\psi}-\mu\psi_{xx}\pm\beta\dot{\psi}=\tilde{\xi}(x,t),\nonumber\\
&&\tilde{\xi}(x,t)=i\sqrt{\frac{\beta\hbar c^3}{2\pi^2
L}}\sum_{n=1}^\infty\int_{-\infty}^{+\infty} \frac{d^3
k}{\sqrt{\omega_{\vec{k}}}}
[b_{n\vec{k}}(0)e^{-i\omega_{\vec{k}}t}-
b_{n\vec{k}}^\dag(0)e^{i\omega_{\vec{k}}t}]\sin\frac{n\pi
x}{L}.\nonumber\\
&&
\end{eqnarray}
By defining
\begin{equation}\label{d3.20}
x_n=\sqrt{\frac{\hbar}{2\lambda\omega_n}}(a_n+a_n^\dag),
\hspace{2.00cm}
p_n=i\sqrt{\frac{\hbar\lambda\omega_n}{2}}(a_n^\dag-a_n),
\end{equation}
 and using (\ref{d3.19}), we obtain
\begin{eqnarray}\label{d3.21}
&&\ddot{x}_n+\omega_n^2x_n+\frac{\beta}{\lambda}\dot{x}_n=\zeta_n(t),\nonumber\\
&&\zeta_n(t)=i\sqrt{\frac{\beta}{4\pi^2\lambda^2}}\int_{-\infty}^{+\infty}\frac{d^3k}
{\sqrt{\omega_{\vec{k}}}}
[b_{n\vec{k}}(0)e^{-i\omega_{\vec{k}}t}-b_{n\vec{k}}^\dag(0)e^{i\omega_{\vec{k}}t}],
\end{eqnarray}
a complete solution of this equation for $t\geq 0$, is
\begin{eqnarray}\label{d3.22}
x_n(t)&=&e^{-\frac{\beta t}{2\lambda}} \{x_n(0)\cos\Omega_n t+
\frac{\beta}{2\lambda\Omega_n}x_n(0)\sin\Omega_n t  -\frac{\beta
M_n(0)}{2\lambda\Omega_n}\sin\Omega_nt\nonumber\\
&-&M_n(0)\cos\Omega_n t
+\frac{\dot{x}_n(0)-\dot{M}_n(0)}{\Omega_n}\sin\Omega_n
t\}+M_n(t),\nonumber\\
 M_n(t)&=&i\int_{-\infty}^{+\infty}d^3 k \sqrt{\frac{\beta \hbar
c^3 }{4\pi^2\lambda^2\omega_{\vec{k
}}}}[\frac{b_{n\vec{k}}(0)}{\omega_n^2-\omega_{\vec{k}}^2-i\frac{\beta}{\lambda}
\omega_{\vec{k}}}e^{-i\omega_{\vec{k}}
t}-\frac{b_{n\vec{k}}^\dag(0)}{\omega_n^2-\omega_{\vec{k}}^2+\frac{i\beta}{\lambda}
\omega_{\vec{k}}}e^{i\omega_{\vec{k}}t}],\nonumber\\
&&
\end{eqnarray}
where $\Omega_n=\sqrt{\omega_n^2-\frac{\beta^2}{4\lambda^2}}$.
This solution consists of two parts, the first part is an
exponentially decreasing function in time which is the solution of
the homogeneous part of the equation (\ref{d3.21}), i.e., when
$\zeta_n(t)=0$, the second part is $M_n(t)$, which is an
oscillatory exponential function and is the response to the noise
force $\zeta_n(t)$. The answer $M_n(t)$, does not have a
classical counterpart and is necessary for a consistent treatment
of dissipative quantum systems. In fact, without $M_n(t)$, the
commutation relations $[x_n,p_n]$, tend to zero in large-time
limit and accordingly, the uncertainty relations would be
violated. From (\ref{d3.91}), (\ref{d3.10}) and (\ref{d3.22}), we
 obtain the asymptotic answer
\begin{eqnarray}\label{d3.23}
&&p_n(t)=\lambda\dot{M}_n(t)+r_N(t)+\beta M_n(t)-\beta
x_n(0),\nonumber\\
&&r_N(t)=\sqrt{\frac{\beta\hbar c^3
}{4\pi^2}}\sum_{n=1}^\infty\int_{-\infty}^{+\infty}\frac{d^3k}{\sqrt{\omega_{\vec{k}}^3}}
[b_{n\vec{k}}(0)e^{-\imath\omega_{\vec{k}}t}+b_{n\vec{k}}^\dag(0)e^{\imath\omega_{\vec{k}}t}],
\end{eqnarray}
for $p_n(t)$, when $t\rightarrow+\infty$.
 If the state of the system in $ t=0 $, is taken to be $
|\psi(0)\rangle=|0\rangle_B\otimes|1_{m_1},...,1_{m_r}\rangle_s $,
where $|1_{m_1},...,1_{m_r}\rangle_s $ is an excited state of the
Hamiltonian $ H_s $, then it is clear that
\begin{equation}\label{d3.24}
\langle\psi(0) \sum_{n=1}^\infty \hbar\omega_n a_n^\dag(0)a_n(0)
|\psi(0)\rangle= \sum_{i=1}^r\hbar\omega_{m_i},
 \end{equation}
  on the other hand, from (\ref{d3.22}), we find
\begin{eqnarray}\label{d3.25}
&&lim_{t\rightarrow\infty}[\langle\psi(0)| : \int_0^L d x[
\frac{1}{2}\lambda\dot{\psi}^2+\frac{1}{2}\mu \psi_x^2]
:|\psi(0)\rangle]\nonumber\\
 &=&\langle\psi(0)| : \sum_{n=1}^\infty
(\frac{1}{2}\lambda
\dot{M}_n^2+\frac{1}{2}\lambda\omega_n^2 M_n^2) :|\psi(0)\rangle= 0.\nonumber\\
&&
\end{eqnarray}
So in the large-time limit, the expectation value of total energy
of the system in the initial state
$|\psi(0)\rangle=|0\rangle_B\otimes|1_{m_1},...,1_{m_r}\rangle_s
$, tends to zero, as expected.

Also, using (\ref{d3.23}), it is clear that\\
 \be\label{d3.26}
lim_{t\rightarrow\infty}[ \langle\psi(0)| : \sum_{n=1}^\infty
\frac{p_n^2}{2\lambda}+\frac{1}{2}\lambda\omega_n^2 x_n^2
:|\psi(0)\rangle] =\frac{\beta^2}{2\lambda}\langle
x_n^2(0)\rangle=\frac{\beta^2\hbar}{4\lambda^2}\sum_{l=1}^r\frac{1}{\omega_{m_r}}.
\ee\\
 Now, if one obtains the Heisenberg equation for  $
b_{n\vec{k}}(t)$ and solve it formally in terms of $
\dot{x}_n(t)$, then using (\ref{d3.22}), he(she) will find\\
\bea\label{d3.27} &&lim_{t\rightarrow\infty}[
\langle\psi(0)|:\int d^3k\sum_{n=1}^\infty
\hbar\omega_{\vec{k}} b_{n\vec{k}}^\dag(t) b_{n\vec{k}}(t):|\psi(0)\rangle]\nonumber\\
&=&\frac{\hbar\beta}{\pi\lambda}\sum_{i=1}^r
\omega_{m_i}\int_0^\infty
\frac{\omega_{mi}^2+x^2}{(\omega_{m_i}^2-x^2)^2+\frac{\beta^2}{\lambda^2}x^2}dx,\nonumber\\
&=&\sum_{=1}^{r}\hbar\omega_{mi}. \eea\\
 Therefore, the energy of the system is completely transferred to the
environment, as expected macroscopically.
\subsection{Transition probabilities}
As in the second section, we can calculate some of transition
probabilities for the scalar field $\psi$ by tracing out the
environment degrees of freedom. For example, Let the total
density operator in the interaction picture at time $t=0$, be
\begin{equation}\label{d3.29}
\rho_I(0)=|r_m\rangle_s\hspace{0.20 cm}_s\langle
r_m|\otimes|0\rangle_B\hspace{0.20 cm}_B \langle 0|,
\end{equation}
where $|0\rangle_B $, is the vacuum state of the environment and $
|r_m\rangle_s=\frac{(a_m^\dag)^r}{\sqrt{r!}}|0\rangle_s $ is an
excited state (Fock state) of the field $\psi$, in large-time
limit, we obtain the following probability transitions
\begin{eqnarray}\label{d3.29}
\Gamma_{|r_m\rangle_s\rightarrow |r_m-1\rangle_s}&=& Tr
[|r_m-1\rangle_s\hspace{0.20
cm}_s\langle r_m-1|\rho(t)],\nonumber\\
&=&Tr_s[|r_m-1\rangle_s\hspace{0.20 cm}_s\langle
r_m-1|\rho_{sI}(t)],\nonumber\\
&=&\frac{4\pi^2 \omega_m^3 r t |f(\omega_m)|^2} {\lambda \hbar
c^3},
\end{eqnarray}
where $Tr_s$, means taking trace over the scalar field
eigenstates.

 Another important case is when the environment has a
canonical thermal distribution
\begin{equation}\label{d3.30}
\rho_I(0)=|r_m\rangle_s\hspace{0.20 cm}_s\langle r_m| \otimes
\rho_B^T, \hspace{2.00cm} \rho_B^T=\frac{e^{\frac{-H_B}{K
T}}}{TR_B(e^{\frac{-H_B}{KT}})},
\end{equation}
then in large-time limit, we obtain the following transition
probabilities
\begin{eqnarray}\label{d35}
&&\Gamma_{|r_m\rangle_s\rightarrow |r_m-1\rangle_s}=
\frac{4\pi^2\omega_m^3
r t|f(\omega_m)|^2 }{\lambda \hbar c^3}\frac{e^{\frac{\omega_m}{K T}}}
{e^{\frac{\omega_m}{K T}}-1},\nonumber\\
&&\Gamma_{|r_m\rangle_s\rightarrow |1_n,r_m\rangle_s}
=\frac{4\pi^2\omega_n^3t}{\lambda \hbar c^3}\frac{
|f(\omega_n)|^2}{ e^{\frac{\omega_n}{K T}}-1},\hspace{1.00
cm}n\neq m,\nonumber\\
&&\Gamma_{|r_m\rangle_s\rightarrow
|r_m+1\rangle_s}=\frac{(r+1)4\pi^2\omega_m^3t}{\lambda \hbar
c^3}\frac{|f(\omega_m)|^2}{ e^{\frac{\omega_m}{K T}}-1}.
\end{eqnarray}
So in low temperatures, the energy flows from the oscillator to
the it's environment by the rate
  $\frac{r\beta}{\lambda}$ and no energy flows from
  the environment to the scalar field.\\
    \section{Dissipative vector field theory}
 Quantization of a quantum vector field $\vec{Y}$ in a three-dimensional
 inhomogeneous absorptive environment can be investigated by modeling the environment
 of $\vec{Y}$ by two independent quantum fields, namely $B$ and $\tilde{B}$ quantum fields.
 The susceptibility of the environment
 and the quantum noise fields are identified in terms of ladder operators of the environment
 and parameters of this model as in previous sections. We assume that the vector field
 $\vec{Y}$ can be propagated in infinite space with a suitable boundary
 condition at infinity, that is, the field $\vec{Y}$ tends to zero at infinity.
 In this case both the fields $B$ and $\tilde{B}$, contain a
 continuum of Klein-Gordon fields. It is also remarkable to note that if the volume
 in which the field $\vec{Y}$ can be propagated
 is a finite volume, for example a cubic cavity, then the fields $B$ and
 $\tilde{B}$, will contain a numerable set of Klein-Gordon fields, as in
 the previous section for the case of a scalar field.

  From the interaction point of view, the field $B$ interacts
 with the conjugate canonical momentum density of the main vector field
 $\vec{Y}$ through a minimal coupling term and quantum field
 $\tilde{B}$ interacts with the field $\vec{Y}$ similar to a dipole
 interaction term. In this scheme, we take the environment Hamiltonian
 as\\
\bea\label{d4.1}
&&H_E=H_B +H_{\tilde{B}},\nonumber\\
&&H_B=\sum_{\nu=1}^3\int d^3\vec{k}\int
 d^3\vec{q}\hbar\omega_{\vec{k}}
b_\nu^\dag(\vec{k},\vec{q},t)b_\nu(\vec{k},\vec{q},t),\nonumber\\
&&H_{\tilde{B}}=\sum_{\nu=1}^3\int d^3\vec{k}\int
d^3\vec{q}\hbar\omega_{\vec{k}}
d_\nu^\dag(\vec{k},\vec{q},t)d_\nu(\vec{k},\vec{q},t). \eea\\
where the annihilation and creation operators  $ b_\nu
(\vec{k},\vec{q},t)$, $ b_\nu^\dag (\vec{k},\vec{q},t)$, $
d_\nu^\dag (\vec{k},\vec{q},t)$ and  $ d_\nu^\dag
(\vec{k},\vec{q},t)$ of the environment, satisfy the commutation
relations
\begin{eqnarray}\label{d4.2}
&&[b_\nu(\vec{k},\vec{q},t),b_{\nu'}^\dag(\vec{k'},\vec{q'},t)]=
\delta_{\nu\nu'}\delta(\vec{k}-\vec{k'})\delta(\vec{q}-\vec{q'}),\nonumber\\
&&[d_\nu(\vec{k},\vec{q},t),d_{\nu'}^\dag(\vec{k'},\vec{q'},t)]=
\delta_{\nu\nu'}\delta(\vec{k}-\vec{k'})\delta(\vec{q}-\vec{q'}).\nonumber\\
\end{eqnarray}
Let us assume that the Hamiltonian of the main system is
\begin{equation}\label{d4.3}
H_Y=\int
d^3\vec{r}[\frac{\vec{\pi}_Y^2}{2\rho}+\frac{1}{2}\rho(\vec{r})\omega_0^2(\vec{r})\vec{Y}^2],
\end{equation}
where $\vec{\pi}_Y$, is the conjugate canonical momentum density
of $\vec{Y}$. An example of this kind of Hamiltonian, the vector
field $\vec{Y}$, can be an electric polarization density in an
absorptive dielectric medium with related eigenfrequency
$\omega_0(\vec{r})$ and density $ \rho(\vec{r})$ \cite{M8}.

 The conjugate fields $\vec{Y}$ and $ \pi_{\vec{Y}}$, satisfy the commutation
relation
\begin{equation}\label{d4.4}
[\vec{Y}_i(\vec{r},t),\vec{\pi}_{Yj}(\vec{r'},t)]=
\imath\hbar\delta_{ij}\delta(\vec{r}-\vec{r'}).
\end{equation}
If the damping forces together with the restoring force
$-\rho(\vec{r})\omega_0^2(\vec{r})\vec{Y}$, are exerted on the
elements of the medium, then the equation of motion of $\vec{Y}$
should be a Langevin-Schr\"{o}dinger equation (\ref{d2.17}),
wherein, the external force $-\vec{\nabla}v$, is replaced with $
-\rho(\vec{r})\omega_0^2(\vec{r})\vec{Y}$. For this purpose, we
take the total Hamiltonian, i.e., the main system (vector field
$\vec{Y}$), plus the environment as
\begin{equation}\label{d4.5}
H=\int d^3\vec{r}[\frac{(\vec{\pi}_Y-\vec{R})^2}{2\rho}+
\frac{1}{2}\rho(\vec{r})\omega_0^2(\vec{r})\vec{Y}^2-\vec{\tilde{R}}\cdot\vec{Y}
+H_B+H_{\tilde{B}}.
\end{equation}
The operators $\vec{R}(\vec{r},t)$ and
$\vec{\tilde{R}}(\vec{r},t)$, play the basic roles in the
interaction between the environment and the system and are
defined by
\begin{eqnarray}\label{d4.6}
\vec{R}(\vec{r},t)&=&\sum_{\nu=1}^3\int d^3\vec{k}\int
\frac{d^3\vec{q}}{\sqrt{(2\pi)^3}}[f(
\omega_{\vec{k}},\vec{r})b_\nu(\vec{k},\vec{q},t)e^{i\vec{q}\cdot\vec{r}}\nonumber\\
&+&f^*( \omega_{\vec{k}},\vec{r})b_\nu^\dag
(\vec{k},\vec{q},t)e^{-i\vec{q}\cdot\vec{r}}]\vec{u}_\nu(\vec{q}),\nonumber\\
\\
 \vec{\tilde{R}}(\vec{r},t)&=&\sum_{\nu=1}^3\int d^3\vec{k}\int
\frac{d^3\vec{q}}{\sqrt{(2\pi)^3}}[g(
\omega_{\vec{k}},\vec{r})d_\nu(\vec{k},\vec{q},t)e^{i\vec{q}\cdot\vec{r}}\nonumber\\
&+&g^*( \omega_{\vec{k}},\vec{r})d_\nu^\dag
(\vec{k},\vec{q},t)e^{-i\vec{q}\cdot\vec{r}}]\vec{u}_\nu(\vec{q}),\nonumber\\
\end{eqnarray}
where
\begin{eqnarray}\label{d34.7}
&&\vec{u}_\nu(\vec{q})=\vec{e}_{\nu\vec{q}},\hspace{2.00cm}\nu=1,2,\nonumber\\
&&\vec{u}_3(\vec{q})=\hat{q}=\frac{\vec{q}}{|\vec{q}|},\nonumber\\
&&\vec{e}_{\nu\vec{q}}\cdot\vec{e}_{\nu'\vec{q}}=\delta_{\nu\nu'},\nonumber\\
&&\hat{q}\cdot\vec{e}_{\nu\vec{q}}=0,\hspace{2.00cm}\nu=1,2,
\end{eqnarray}
 are three orthogonal unit vectors for any $\vec{q}$.

 The function $f(\omega_{\vec{k}},\vec{r})$ and
 $f(\omega_{\vec{k}},\vec{r})$,
 are the coupling functions, which are position dependent(independent)
for an inhomogeneous(homogeneous) environment respectively. The
equations of motion for the fields $\vec{Y}$ and $\vec{\pi}_Y $,
can be obtained from the Heisenberg equations
\begin{eqnarray}\label{d4.8}
&&\frac{\partial\vec{Y}}{\partial
t}=\frac{i}{\hbar}[H,\vec{Y}]=\frac{\vec{\pi}_Y-\vec{R}}{\rho},\nonumber\\
&&\frac{\partial\vec{\pi}_Y}{\partial
t}=\frac{i}{\hbar}[H,\vec{\pi}_Y]=-\rho\omega_0^2(\vec{r})\vec{Y}+\vec{\tilde{R}}.
\end{eqnarray}
By eliminating $\vec{\pi}_Y$ between these equations we obtain
the equation of motion of the vector field $\vec{Y}$ as,
\begin{equation}\label{d4.9}
\rho\frac{\partial^2\vec{Y}}{\partial
t^2}+\rho\omega_0^2(\vec{r})\vec{Y}=-\frac{\partial\vec{R}}{\partial
t}+\vec{\tilde{R}}.
\end{equation}
Similar to what we did in the previous sections one can obtain,
 the constitutive equation of the environment as
\begin{eqnarray}\label{d4.12}
&&\vec{R}(\vec{r},t)=\vec{R}_N(\vec{r},t)+\int_0^{|t|}
dt'\chi(\vec{r},|t|-t')\dot{\vec{Y}}(\vec{r},\pm t'),\nonumber\\
&&\vec{\tilde{R}}(\vec{r},t)=\vec{\tilde{R}}_N(\vec{r},t)+\int_0^{|t|}
dt'\tilde{\chi}(\vec{r},|t|-t')\vec{Y}(\vec{r},\pm t'),
\end{eqnarray}
where the upper(lower) sign corresponds to $t>0$$(t<0)$,
respectively. The following relations
\begin{eqnarray}\label{d4.13}
&&\chi(\vec{r},t)=\frac{8\pi}{\hbar c^3}\int_0^\infty d \omega
\omega^2|f(\omega,\vec{r})|^2\sin\omega t \hspace{1.50cm}t>0\nonumber\\
&&\chi(\vec{r},t)=0\hspace{6.50cm}t\leq0
\end{eqnarray}
\begin{eqnarray}\label{d4.14}
&&\tilde{\chi}(\vec{r},t)=\frac{8\pi}{\hbar c^3}\int_0^\infty d
\omega
\omega^2|g(\omega,\vec{r})|^2\sin\omega t, \hspace{1.50cm}t>0,\nonumber\\
&&\tilde{\chi}(\vec{r},t)=0,\hspace{6.50cm}t\leq0,
\end{eqnarray}
give the susceptibilities of the environment in terms of the
coupling functions. Operators $\vec{R}_N$ and $\vec{\tilde{R}}_N$,
are noise fields
\begin{eqnarray}\label{d4.15}
\vec{R}_N(\vec{r},t)&=&\sum_{\nu=1}^3\int d^3\vec{k}\int
\frac{d^3\vec{q}}{\sqrt{(2\pi)^3}}[f(\omega_{\vec{k}},\vec{r})b_\nu(\vec{k},\vec{q},0)
e^{-\imath\omega_{\vec{k}}t+\imath\vec{q}\cdot\vec{r}}\nonumber\\
&+&f^*( \omega_{\vec{k}},\vec{r})b_\nu^\dag(\vec{k},\vec{q},0)
e^{\imath\omega_{\vec{k}}t-i\vec{q}\cdot\vec{r}}]\vec{u}_\nu(\vec{q}),\nonumber\\
\\
\vec{\tilde{R}}_N(\vec{r},t)&=&\sum_{\nu=1}^3\int d^3\vec{k}\int
\frac{d^3\vec{q}}{\sqrt{(2\pi)^3}}[g(
\omega_{\vec{k}},\vec{r})d_\nu(\vec{k},\vec{q},0)e^{-\imath\omega_{\vec{k}}t+
\imath\vec{q}\cdot\vec{r}}\nonumber\\
&+&g^*( \omega_{\vec{k}},\vec{r})d_\nu^\dag
(\vec{k},\vec{q},0)e^{\imath\omega_{\vec{k}}t-\imath\vec{q}\cdot\vec{r}}]
\vec{u}_\nu(\vec{q}).\nonumber\\
\end{eqnarray}
Finally, substituting (\ref{d4.12}) in (\ref{d4.9}), we find a
 generalized Langevin-Shr\"{o}dinger equation for the damped
vector field $\vec{Y}$
\begin{eqnarray}\label{d4.16}
&&\rho\ddot{\vec{Y}}+\rho\omega_0^2(\vec{r})\vec{Y}\pm
\frac{d}{dt}\int_0^{|t|}dt'\chi(\vec{r},|t|-t')
\dot{\vec{Y}}(\vec{r},\pm t')-\nonumber\\
&&\int_0^{|t |}dt'\tilde{\chi}(\vec{r},|t|-t')\vec{Y}(\vec{r},\pm
t')+\dot{\vec{R}}_N-\vec{\tilde{R}}_N\vec{\xi}(\vec{r},t)=0,
\end{eqnarray}

 This equation can be solved using the Laplace transformation technique
for negative and positive times similar to the previous sections.
\section{Concluding remarks}
By considering a dissipative quantum system as an open system
which interacts with it's environment, a modeling of the
environment by massless Klein-Gordon fields, is achieved. Using
the above idea, i.e., the field version of an environment, an
arbitrary dissipative quantum system is quantized systematically
and consistently. Some coupling functions are introduced which
physically describe the environment under consideration and also
have a basic role in interaction between the system and it's
environment. Inspired by the ideas of electrodynamics in a media,
some susceptibility functions are attributed to the environment
and formulas connecting these susceptibilities to the coupling
function, are found. The quantum Langevin-Schr\"{o}dinger equation
is obtained directly from the Heisenberg equations. The explicit
form of the noise terms are obtained. Some transition
probabilities indicating the way energy flows from the system to
it's environment, are calculated and the energy conservation, as a
signal of consistency, is explicitly worked out. The whole
formalism is generalized to the case of a dissipative scalar and
vector field theory straightforwardly.

\end{document}